\documentclass[sigconf,nonacm]{acmart}
\AtBeginDocument{%
  }

\usepackage{graphicx}
\setkeys{Gin}{draft=false}

\usepackage{amsmath}
\usepackage{bm}

\usepackage{booktabs}
\usepackage{tabularx}
\usepackage{array}

\usepackage{xcolor}
\newif\ifshowqs
\showqsfalse 

\definecolor{qscolor}{RGB}{0,120,60}
\newcommand{\Q}[1]{%
  \ifshowqs
    \par\noindent
    {\textcolor{qscolor}{\textbf{[Q: #1]}}}\par\noindent
  \fi
}
\begin{document}

\title{What Does It Take to Get Guarantees? \\Systematizing Assumptions in Cyber-Physical Systems}

\author{Chengyu Li}
\affiliation{%
  \institution{University of Florida}
  \city{Gainesville}
  \country{USA}}
\email{lichengyu@ufl.edu}

\author{Saleh Faghfoorian}
\affiliation{%
  \institution{University of Florida}
  \city{Gainesville}
  \country{USA}}
\email{faghfoorian.m@ufl.edu}

\author{Ivan Ruchkin}
\affiliation{%
  \institution{University of Florida}
  \city{Gainesville}
  \country{USA}}
\email{iruchkin@ece.ufl.edu}

\begin{abstract}
Formal guarantees for cyber-physical systems (CPS) rely on diverse assumptions. If satisfied, these assumptions enable the transfer of abstract guarantees into real-world assurances about the deployed CPS. Although assumptions are central to assured CPS, there is little systematic knowledge about what assumptions are made, what guarantees they support, and what it would take to specify them precisely. To fill this gap, we present a survey of assumptions and guarantees in the control, verification, and runtime assurance areas of CPS literature. From 104 papers over a 10-year span (2014–2024), we extracted 423 assumptions and 321 guarantees using grounded-theory coding. We also annotated the assumptions with 21 tags indicating elementary language features needed for specifications. Our analysis highlighted prevalent trends and gaps in CPS assumptions, particularly related to initialization, sensing, perception, neural components, and uncertainty. Our observations culminated in a call to action on reporting and testing CPS assumptions. 

\end{abstract}




\maketitle

\section{Introduction}
\looseness=-1
Ensuring that a cyber-physical system (CPS) satisfies its desired properties has been a significant research theme. A \emph{guarantee} is a falsifiable property of a deployed CPS operating in its environment, derived from a theoretical or computational analysis (e.g., a theorem or a model checking tool)~\cite{bliudze_rigorous_2017, kress-gazit_formalizing_2021}. The properties of interest include safety, security, stability, performance, robustness, and others. 

An \emph{assumption} is a necessary condition, presupposed in the analysis process, for a guarantee to hold~\cite{broy_theory_2017, bulandran_exploration_2012}. If all required assumptions are satisfied for a particular CPS deployment, the abstract guarantee claim turns into a real-world fact. Thus, conceptually, an assumption serves as the ``bridge'' between what is proven mathematically/computationally and what should hold for the deployed system. Specifying assumptions is a key engineering task that requires domain expertise. If assumptions fail (e.g., due to biased training data, invalid abstractions, or rare events), the respective guarantees should not be trusted. As a result, assumptions play a key role in system assurance~\cite{koopman_how_2022, carr_quantifying_2023, yang_assumptions_2018}.

Due to the complexity of modern CPS, assumptions come in a dizzying \textit{variety} of forms and granularities. They can be made about software/hardware components, external environments, or data distributions. In the context of sensing, assumptions can concern sensor coverage, latency, update rate, calibration, and measurement noise. Perception assumptions can relate to observability, filter stability, and outlier handling, while assumptions about dynamics and actuation can determine model fidelity, traction, and actuator limits. The assumptions on the external environments can constrain disturbance envelopes, other agents’ behavior models, and possible faults. Hence, the sheer diversity of assumptions makes it difficult to handle them systematically.

Despite their major role, CPS assumptions are \textit{rarely specified} in a precise, unified, and testable form. In practice, most assumptions are informal, fragmented, and come from heterogeneous reasoning styles (e.g., deterministic vs. probabilistic)~\cite{fu_framework_2020, fu_uacfinder_2020}. Most commonly, assumptions are formalized in assume-guarantee frameworks~\cite{sokolsky_monitoring_2016, hutchison_assume-guarantee_2012, li_stochastic_2017, incer_pacti_2023}, which impose a particular syntax and semantics to empower compositional reasoning, treating assumptions as formally provable statements --- rather than interfaces to the inherently informal world. However, recently there has been an increasing interest in assumptions for validation~\cite{dsouza_towards_2026, ferrando_recognising_2018}, monitoring~\cite{deshmukh_monitorability_2020, ruchkin_confidence_2022, feng_assumption-based_2021}, and control synthesis~\cite{meng_automated_2024, carr_quantifying_2023, anand_computing_2023}.

\looseness=-1
The CPS community faces \textit{two immediate challenges} in mastering its assumptions. First, we lack a clear taxonomy of \textit{what} assumptions are actually being made, and \textit{which} guarantees they are intended to support. Second, our efforts in specifying assumptions are not guided by the relevant \textit{language features}, such as probabilistic, temporal, and neural predicates. These obstacles hinder the development of successful specification, management, and validation frameworks for CPS assumptions.  Unsurprisingly, existing assumption specifications~\cite{de_saqui-sannes_making_2016, mohammadinejad_mining_2020, finkbeiner_introspective_2019} are insufficiently expressive for CPS.

This paper puts forward a \textit{literature survey} to address these two challenges, filling the gap in our understanding of modern CPS assumptions in control, verification, and runtime assurance. The related surveys have been primarily targeting two niches: (i) assumptions in software engineering~\cite{yang_assumptions_2018, paech_classifying_2008, bulandran_exploration_2012}, which ignored the physical aspects and diverse guarantees, and (ii) modeling for CPS~\cite{barisic_multi-paradigm_2022, graja_comprehensive_2018, barisic_modelling_2025}, which ignored the role and richness of assumptions. In contrast, our survey formulates and answers \textit{three research questions:}
\begin{enumerate}
    \item What assumptions and guarantees are asserted for CPS? 
    \item What types of assumptions support which guarantees?
   \item What language elements would formalize assumptions?
\end{enumerate}

To answer these questions, we collected a corpus of 104 CPS papers (2014--2024) from control (36), verification (43), and runtime assurance (25). We extracted a total of 423 distinct assumptions and 321 guarantees. Following the grounded theory method~\cite{corbin_basics_2015, glaser_discovery_1999}, we coded assumptions with 14 tags and grouped them into 4 high-level categories (Physical, Modeling, Interface/Estimation, External Constraints), whereas the guarantees were labeled with 9 tags. For every assumption, we attempted a minimal formalization and tagged it with at least one of 21 categories of language features 
 (e.g., variables, quantifiers, algebraic relations, probabilistic statements, temporal operators, neural predicates). 

Finally, we extracted insights and literature gaps from our annotated data. Our analysis revealed \textit{several broad patterns}: model-related assumptions are most prevalent; safety is the most common guarantee but is often limited by initialization conditions; sensing and perception assumptions are frequently underspecified; neural assumptions are rarely stated explicitly; and assumption statements tend to rely mainly on first-order algebraic forms, with some use of uncertainty. Based on these observations, we put forth \textit{four suggestions} for the CPS community: (a) make initialization dependence explicit, (b) describe and test sensing and perception constraints, (c) get more precise about the assumptions on neural architecture and behavior, and (d) report stochastic uncertainty in a structured way. For transparency, we release our \href{https://docs.google.com/spreadsheets/d/e/2PACX-1vRd6uMD7AQFMFZUouHEKP923pCrlstxin7xkRuTwRGkfXJ2S6dByGgaXCpOUNWd4UAqsv1UB9fU0kvH/pubhtml}{\textbf{tagged database}}. Furthermore, to continually improve our systematization of assumptions, we invite the CPS community to provide pointers to overlooked studies and assumptions via an \href{https://docs.google.com/forms/d/e/1FAIpQLSfuPPUTjk_LMuJDBtHnVeN2szL1k7KP9pccBik7Q14lMJ4ksw/viewform?usp=dialog}{\textbf{online form}}.

The remainder of the paper is organized as follows. Section~\ref{sec:survey-methodology} defines the survey scope, describes the literature corpus, and details the grounded-theory coding procedure (open, axial, and selective), together with the assumption/guarantee codebooks and the language-feature taxonomy. Section~\ref{sec:summary-of-literature-data} then reports descriptive summaries, including the distributions of assumption categories and tags, guarantee types, language features, and assumption–guarantee co-occurrence patterns. Section~\ref{sec:high-level-observations} interprets these findings with higher-level observations, building on which, Section~\ref{sec:calls-for-action} proposes four assumption reporting guidelines for CPS.
Finally, Section~\ref{sec:limitations} discusses the limitations of our study.
\section{Survey Methodology}
\label{sec:survey-methodology}

This section explains how we construct and analyze our corpus of assumption and guarantee statements in CPS. We first define the scope of the survey, along with brief formal definitions of assumptions and guarantees. We then describe how we selected 104 CPS papers published between 2014 and 2024. The remainder of the section details our three-step process of annotating the paper data: (i) \textit{open coding} to identify assumptions, guarantees, violation consequences, and language features; (ii) \textit{axial coding} to assign multi-label tags to each; and (iii) \textit{selective coding} to organize these tags into higher-level categories.

\subsection{Survey Scope}
\label{subsec:goal-scope}
\Q{Which CPS components (sensing → control → environment) are in scope?}

\subsubsection{CPS Components in Scope}
We scope the survey to the full sensing–decision–actuation–environment loop that defines a closed-loop CPS. In the context of this loop, we focus our paper selection on three technical areas: \emph{control}, \emph{verification}, and \emph{runtime assurance}.  

Control papers address the synthesis and analysis of controllers and planners that operate under model uncertainty, distributional shift, or safety constraints. Verification papers focus on formal analysis and proof techniques that establish correctness or robustness properties of these controllers and their learned components. Runtime assurance papers study online validation, enforcement, and recovery mechanisms that ensure guarantees continue to hold during deployment. From offline design to real-world operation, these technical areas capture how assumptions and guarantees are defined, verified, and maintained. By practical necessity, our focus meant that we had to exclude other active areas, such as human-CPS interaction, real-time scheduling, and CPS security. 

\Q{What is an assumption? What is a guarantee?}
\subsubsection{Definitions of Assumptions and Guarantees}
First, we introduce four abstract sets that pertain to a CPS. Let $\mathcal{W}$ denote the set of potential real-world systems, $\mathcal{M}$ the set of corresponding abstractions or models, $\mathcal{X}$ the set of \emph{offline} artifacts used during design or verification (e.g., training datasets, calibration logs, or pre-trained neural modules), and $\mathcal{E}$ the set of \emph{online} environment conditions encountered during operation (e.g., road curvature, lighting, weather, network latency, or human behavior). These sets constitute a universe of discourse:

\begin{definition} [Universe of Discourse]
\label{def:universe}
The universe of discourse 
\[
\mathcal{U} = (\mathcal{W}\times\mathcal{E}) \times (\mathcal{M}\times\mathcal{X})
\]
is the set of all configurations $u = (w,e,m,x)$ that pair a real system and environment $(w,e)$ with a design-time abstraction and artifacts $(m,x)$.
\end{definition}

\begin{definition}[Abstract Guarantee]
\label{def:abstract-guarantee}
An abstract guarantee $\hat{G}$ is a binary-valued property
\[
\hat{G} : \mathcal{M} \times \mathcal{X} \rightarrow \{\top,\bot\},
\]
defined over models and their associated offline artifacts. For any
$(m,x) \in \mathcal{M} \times \mathcal{X}$, ``$(m,x) \models \hat{G}$'' indicates that the abstract model
$m$ together with artifacts $x$ satisfies $\hat{G}$. Such guarantees
can be established purely at the model level (e.g., via theorem proving
or reachability analysis), without tying them yet to any particular
physical system or environment.
\end{definition}

\begin{definition}[Operational Guarantee]
\label{def:operational-guarantee}
An operational guarantee $G$ is a falsifiable statement 
\[
G : \mathcal{W} \times \mathcal{E} \rightarrow \{\top, \bot\},
\]
defined over the set of possible real systems $\mathcal{W}$ and their operating environments $\mathcal{E}$. For any configuration $(w,e) \in (\mathcal{W} \times \mathcal{E})$, the statement ``$w,e \models G$'' means that the concrete system instance $w$ satisfies the claimed property $G$ when operating under environmental condition $e$.
\end{definition}

Now that we drew a distinction between abstract guarantees $\hat{G}$ and operational guarantees $G$, what is the connection between the two? This is where assumptions come in. 

\begin{definition}[Assumption]
\label{def:assumption}
An assumption is a predicate 
\[
A : \mathcal{U} \rightarrow \{\top, \bot \}
\]
that is both falsifiable and satisfiable over a universe of discourse $\mathcal{U}$. For any configuration $u = (w,e,m,x) \in \mathcal{U}$, ``$u \models A$'' means the assumption holds for that particular pairing of design-time abstraction and real-world execution. 
\end{definition}

The point of an assumption is to specify the conditions under which an abstract guarantee $\hat{G}$ established on $(m,x)$ can be trusted to apply to the operational world $(w,e)$ as some operational guarantee $G$. An example that spans all four components $(w,e,m,x)$ is the ``no distribution shift'' assumption~\cite{cairoli_neural_2021}: the runtime observations generated by the real system and its environment $(w,e)$ follow the same underlying data distribution as the offline dataset $x$ used to train and calibrate the model $m$. This statement is satisfiable (e.g., under an i.i.d. sampling process) and falsifiable (e.g., via tests for distribution shift). When this assumption is violated, finite-sample validity and safety confidence provided by conformal-prediction monitoring can no longer be trusted~\cite{cairoli_neural_2021}.

Here is how we formalize the relationship between assumptions, abstract guarantees, and operational guarantees. For a configuration $u = (w,e,m,x)$, the assumption $A(u)$ connects the real-world pair $(w,e)$ with its abstraction $(m,x)$ and specifies when an abstract guarantee can be transferred to an operational guarantee on the physical system. Thus, the relationship between the assumptions, abstract guarantees, and operational guarantees is:
\begin{equation}
\label{eq:a2g}
\bigl(m,x \models \hat{G}\bigr) \land (u \models A)
\;\Rightarrow\;
\bigl(w,e \models G\bigr).
\end{equation}

This formulation highlights the major role of assumptions in assuring real-world CPS: if assumptions are satisfied, guarantees apply to the real world; otherwise, they remain abstract and disconnected from the reality. In the remainder of the paper, we use the term ``guarantee'' to denote an abstract guarantee $\hat{G}(m,x)$, unless explicitly stated otherwise for notational convenience.

\Q{What is the general structure of this survey?}
\subsubsection{Survey Structure and Coding Methodology}
We applied the \textit{grounded theory methodology} because it supports qualitative analysis in emerging research areas where prior concepts are limited and predefined categories are unavailable~\cite{myers_qualitative_2013, lazar_research_2017}. Following the classical approach by Corbin and Strauss~\cite{corbin_basics_2015}, we used a three-step coding (i.e., data annotation) process to extract and interpret assumptions and guarantees. In \emph{open coding}~\cite{corbin_basics_2015}, we identified and labeled quotations describing guarantees, the implicit and explicit assumptions that support them, and any statements about the effect on guarantees if the assumption does not hold (which helps determine which guarantees depend on which assumptions). In \emph{axial coding}~\cite{lyberg_survey_1997, preece_interaction_2015, corbin_basics_2015}, we refined these initial labels into stable \emph{tag} codebooks for assumptions, guarantees, and language features through iterative comparison and merging. In \emph{selective coding}~\cite{corbin_basics_2015, kuziemsky_grounded_2007}, we integrated the axial-level assumption tags into a small set of higher-level categories and constructed the assumption\,$\times$\,guarantee tables used in our quantitative analysis, from which we derived the corpus-level patterns reported in Sec.~\ref{sec:summary-of-literature-data}. Figure~\ref{fig:coding-workflow} summarizes this three-step workflow and the resulting artifacts.

\begin{figure}[bth]
  \centering
  \includegraphics[width=\linewidth]{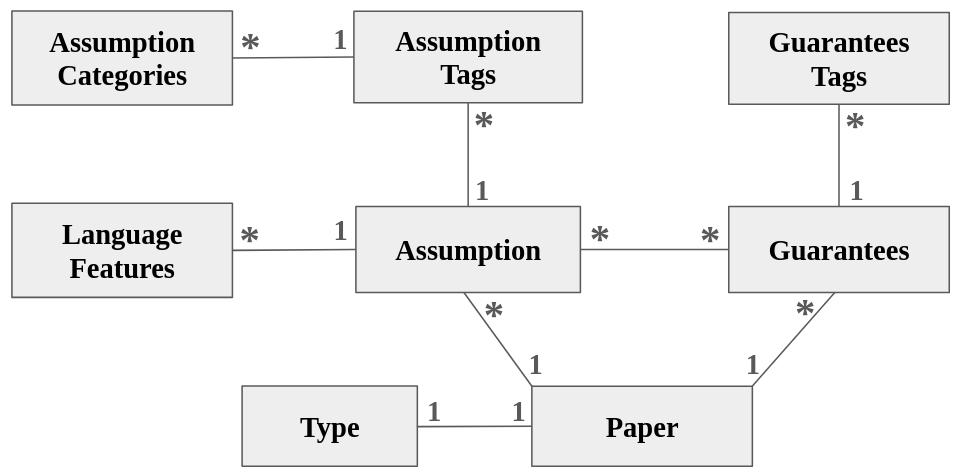}
  \vspace{-7mm}
  \caption{Key concepts of our survey and their relationships;  1 means a ``single concept'', {\Large$\ast$} means ``one or more concepts''. }
  \vspace{-3mm}
  \Description{Entity–relationship diagram showing how survey data is organized across entities such as questions, responses, and participants.}
  \label{fig:overall-erd}
\end{figure}

Fig.~\ref{fig:overall-erd} summarizes how the concepts in this survey are related. Each paper contributes one or more extracted assumptions and guarantees. Every assumption and guarantee receives one or more tags that are subsequently mapped to higher-level categories for analysis. Each assumption is also annotated with the language features in its attempted formalization. Papers are labeled by their main technical area: control, verification, or runtime assurance.

\subsection{Corpus}
\begin{figure}[h]
    \centering
    \includegraphics[width=1\linewidth]{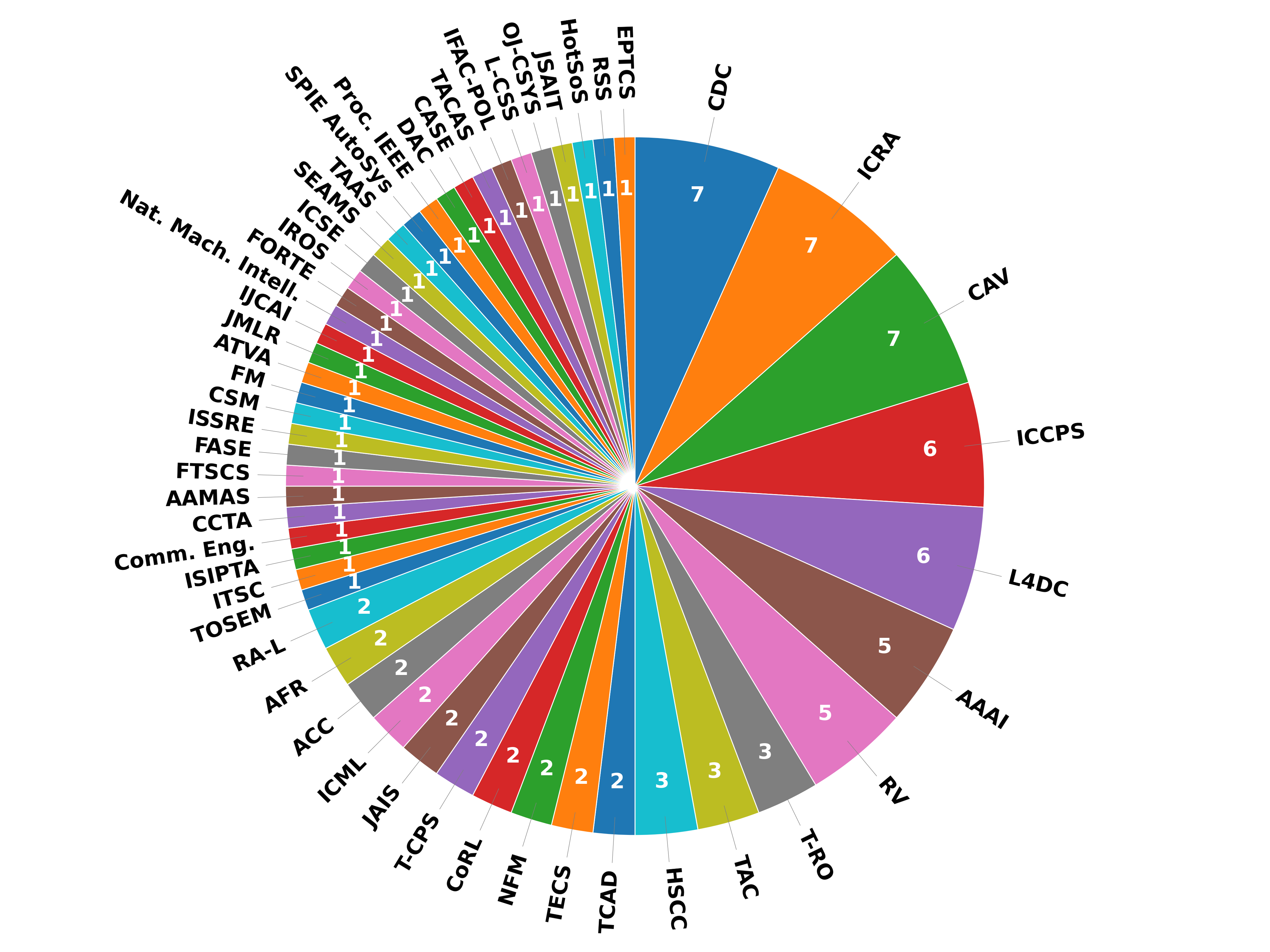}
    \vspace{-7mm}
    \caption{Publication venues of the 104 surveyed papers.}
    \vspace{-4mm}
    \label{fig:pie_venues}
\end{figure}

\Q{When collecting papers for your survey, which research conferences/journals and which time period are you considering?}
Our final dataset consists of 423 assumptions and 321 guarantees extracted from 104 papers published between 2014 and 2024. These papers span three CPS areas: verification ($43/104$), control ($36/104$), and runtime assurance ($25/104$). The most represented venues are CDC, ICRA, and CAV (7 papers each), followed by L4DC and ICCPS (6 each), and then AAAI and RV (5 each). Fig.~\ref{fig:pie_venues} summarizes the distribution of papers across venues, showing that our corpus is drawn from a diverse set of key CPS venues. To diversify our corpus within practical limits, we excluded papers without obvious guarantees and with contributions similar to those already present in our corpus (e.g., extended versions). 

\subsection{Coding Process}
\begin{figure}[t]
  \centering
  \includegraphics[width=\linewidth]{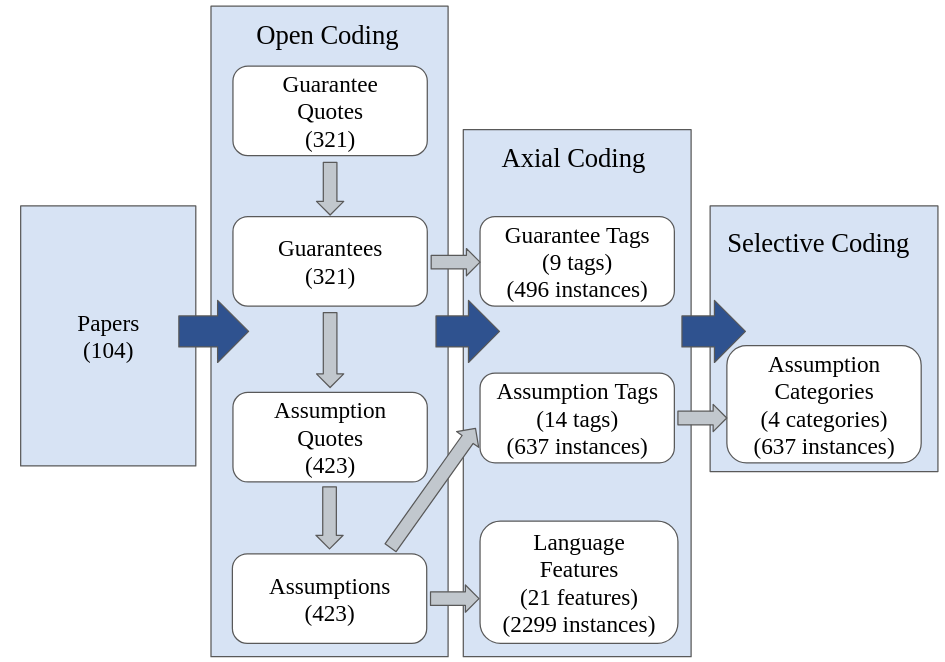}
  \vspace{-7mm}
  \caption{Survey workflow from papers to analysis.}
  \label{fig:coding-workflow}
\end{figure}

\Q{How were assumptions and guarantees identified?}
We conducted the data annotation in three steps illustrated in Fig.~\ref{fig:coding-workflow}: open coding to identify assumptions and guarantees in each paper, axial coding to organize these findings into consistent tags, and selective coding to group these tags into higher-level categories for comparison and trend analysis.

\subsubsection{Open Coding: Identifying Assumptions and Guarantees}

For each paper, we conducted inductive open coding to extract (i) \emph{guarantees} --- explicit formal claims about the system behavior or performance; (ii) one or more supporting \emph{assumptions} --- necessary conditions on world, model, data, timing, or components under which the guarantee is claimed; and (iii) the \emph{assumption-violation consequence} --- a statement describing what happens if the assumption does not hold. We treated common lexical cues (``assume,'' ``under,'' ``given,'' ``subject to,'' ``provided that,'' ``holds when,'' ``if\ldots then\ldots'') as inclusive triggers to avoid bias toward the single token ``assume''. We also allowed multi-sentence spans when required to preserve the intended semantics. These assumptions and guarantees were organized in a structured table, where each row represents a distinct assumption instance (Assumption-ID) extracted from a specific paper. For each assumption, the table contained the corresponding bibliographic metadata (title, authors, venue, and year), the brief paraphrased statement, its formalized representation, the associated language features, the linked Guarantee-ID(s) and the paraphrased guarantee(s), the stated violation consequences, and three source quotation fields referencing the assumption, the guarantee, and the described consequences of assumption violations in the original text. We excluded trivial and non-falsifiable assumptions, such as ``we assume all variables are well-defined'' or ``we assume the environment behaves reasonably'' as they neither constrain the universe of discourse nor admit a meaningful notion of violation.

\Q{How were assumptions and guarantees tagged?}

\subsubsection{Axial Coding: Assigning Tags and Language Features}
We grouped assumption and guarantee quotes using multi-label axial coding~\cite{corbin_basics_2015}. We focused assumption tags on the CPS-specific meaning of an assumption and its role in the closed loop (e.g., plant physics, environment/agents, sensing, perception, timing, communication, control/actuation, neural components). The resulting tags are shown in Tab.~\ref{tab:assumption-tags}.

This codebook was developed by iterative comparison and refinement: seed labels were proposed from a pilot batch, iteratively merged or split until definitions were non-overlapping and exhausted all instances, then frozen for corpus-wide tagging. Tagging was applied at the quote level and could assign multiple tags per instance (e.g., a timing constraint on a perception pipeline is tagged \texttt{A\_TIME} and \texttt{A\_PERCEP}). To resolve common ambiguities, we treated \texttt{A\_PHY} as claims about real-world physics independent of a particular model, \texttt{A\_ABS} as properties of the chosen representation/abstraction, and \texttt{A\_MVAL} as claims about model fidelity or domain of validity.

\begin{table}[t]
  \centering
  \footnotesize
  \setlength{\tabcolsep}{4pt}
  \renewcommand{\arraystretch}{1.05}
  \caption{Assumptions codebook from axial coding.}
  \vspace{-2.5mm}
  \label{tab:assumption-tags}
  \begin{tabular}{p{0.4\linewidth} p{0.58\linewidth}}
    \toprule
    \textbf{(Tag) Name} & \textbf{Meaning} \\
    \midrule
    (A\_PHY) Physical dynamics & Real‑world physics (continuous/hybrid) governing state evolution prior to any abstraction. \\
    (A\_ENV) Environment \& agents & Exogenous world/agents that perturb or constrain the system. \\
    (A\_HUM) Human behavior & Models of human behavior/compliance when people are in the loop. \\
    (A\_ABS) System abstraction & Properties of the chosen mathematical representation/structure enabling analysis. \\
    (A\_MVAL) Model validity & Fidelity of the model to the plant over the stated operating domain. \\
    (A\_INIT) State envelope & Declared initial/operational set/tube within which the claims hold. \\
    (A\_ACTU) Actuation feasibility & Physical/solver limits and feasibility of control actions (rates, saturation). \\
    (A\_SENS) Sensing & Sensor/estimator latency, noise, calibration, observability, and error bounds. \\
    (A\_PERCEP) Perception & Reliability of learned perception/detection interpreting the environment. \\
    (A\_TIME) Timing \& scheduling & Compute/comm. sampling, deadlines, jitter, and scheduling constraints. \\
    (A\_NN) Neural regularity & Regularity/calibration/robustness bounds on learned mappings (e.g., Lipschitz). \\
    (A\_ARCH) Neural architecture & Architectural and interface constraints of learned components. \\
    (A\_FAULT) Fault \& tamper model & Assumed fault/attack/spoof/dropout classes and their rates. \\
    (A\_UTIL) Utility components \& priors & Fixed tools/priors the system relies on (simulators, cost maps, embeddings). \\
    \bottomrule
  \end{tabular}
\end{table}

\Q{How do we find out what language tools are needed to describe assumptions?}
In parallel with these tags, we also performed axial coding on the formal symbolic expression of each assumption. For every assumption quote, we paraphrased it into a falsifiable statement, attempted a symbolic formalization, and recorded the minimal syntactic \emph{language features} needed to express that formalization. The resulting language-feature codebook in Tab.~\ref{tab:language-features} enumerates recurring building blocks: sets, arithmetic, logic, quantifiers, temporal operators, probability, distributions, dynamics, matrices, state machines, counters, algorithms, and neural components. Whereas assumption tags capture \emph{what} an assumption asserts about a CPS (e.g., environment, sensing, model validity), language features capture \emph{how} that condition is written down. Both codebooks are applied at the quote level and may assign multiple entries per instance.

\begin{table}[t]
  \centering
  \footnotesize
  \setlength{\tabcolsep}{4pt}
  \renewcommand{\arraystretch}{0.98}
  \caption{Language features for expressing assumptions from open coding.}
  \vspace{-2.5mm}
  \label{tab:language-features}
  \begin{tabularx}{\linewidth}{@{}l X@{}}
    \toprule
    \textbf{Feature} & \textbf{Meaning} \\
    \midrule
    State variables & Symbols denoting system state, inputs, outputs, parameters, indices, or time in the domain of discourse. \\
    Arithmetic & Algebraic expressions, assignments, and (in)equality relations over numeric domains. \\
    Boolean & Logical connectives and implication with truth‑valued predicates (e.g., $\land,\lor,\lnot,\to$). \\
    Set operations & Membership and set algebra ($\in,\subseteq,\cup,\cap$), Cartesian products, images/pre‑images, and containment relations. \\
    Norms & Metric quantities ($\lVert\cdot\rVert$) and norm‑bounded relations (e.g., Lipschitzness, distance constraints). \\
    Matrices & Linear‑algebraic objects and properties (products, transpose/inverse, rank/eigenvalues, positive‑(semi)definiteness, LMIs). \\
    Set aggregations & Aggregators over collections or windows (sum, count, integrals, minima/maxima, suprema/infima). \\
    System components & References to named system entities (models/blocks such as dynamics $f$, controllers, generators, partitions, signals). \\
    State machines & Discrete or hybrid automata with modes/states, transitions, guards, resets, and invariants. \\
    Derivatives & Differential/integro‑differential operators and ODE/PDE expressions describing temporal evolution. \\
    First‑order quantifiers & Universal/existential quantification over individuals (states, inputs, indices, time). \\
    Linear temporal logic (LTL) & Temporal modalities over linear traces (e.g., $\Box$, $\Diamond$, $U$, $X$) expressing safety/liveness over time. \\
    Path quantifier (CTL) & Branching‑time path quantification (A/E) over executions or schedulers. \\
    Counters & Cardinality bounds on events within horizons or sliding windows (e.g., weakly‑hard patterns). \\
    Metric/signal temporal logic & Real‑time temporal logics with explicit time intervals; STL admits quantitative robustness semantics. \\
    Second‑order quantifiers & Quantification over functions, policies, or controllers (e.g., $\exists\pi$, $\forall f$). \\
    Probability & Probability statements and chance‑constraint thresholds on events. \\
    Statistical & Statistical functionals/estimators and performance metrics (expectation, variance, quantiles, sample‑based measures). \\
    Distribution & Assumptions on probability distributions (i.i.d., independence, family/parameters, bounded density, moments). \\
    Neural & Predicates referencing learned modules (perception/policy/generative) as system components. \\
    Algorithm & Imperative/procedural constructs for monitoring, decision logic, or control switching. \\
    \bottomrule
  \end{tabularx}
\end{table}

Finally, we developed a \emph{guarantee} codebook (shown in Tab.~\ref{tab:guarantee-tags}) that labels each claim about the system's desired property (e.g., safety, stability, robustness). As with assumptions, tags were created and refined by inductive axial coding with iterative comparison until definitions stabilized. Tagging was multi-label and applied at the quote level, and we did not impose exclusivity or precedence among tags. We tagged guarantees by their intended CPS impact, rather than theorem names or tool outputs. For traceability, we recorded the detailed formal subtype (e.g., barrier invariance, Lyapunov/Input-to-State Stability (ISS), chance constraints) in a separate field.

When a guarantee claim spanned multiple aspects, all relevant tags were applied; for instance, ``probabilistic safety at horizon $T$'' received \texttt{G\_SAFE} and \texttt{G\_ROBUST}; ``existence of a feasible controller under timing constraints'' received \texttt{G\_FEAS} and \texttt{G\_TIME}. Ambiguous cases were provisionally multi-tagged and later adjudicated during selective coding by inspecting the formal subtype and local context; for instance, barrier invariance $\rightarrow$ \texttt{G\_SAFE}; Lyapunov or Input-to-State Stability (ISS) $\rightarrow$ \texttt{G\_STAB}; chance constraints $\rightarrow$ \texttt{G\_ROBUST}; availability/mean time between failures $\rightarrow$ \texttt{G\_RELY}; recovery after faults $\rightarrow$ \texttt{G\_RESIL}; deadlines/jitter $\rightarrow$ \texttt{G\_TIME}; controller/plan existence $\rightarrow$ \texttt{G\_FEAS}; objective accuracy/optimality $\rightarrow$ \texttt{G\_PERF}; adversary/tamper/CIA $\rightarrow$ \texttt{G\_SECURE}. The codebook was frozen once new instances mapped cleanly to existing tags without requiring additional splits.

\begin{table}[t]
  \centering
  \footnotesize
  \setlength{\tabcolsep}{4pt}
  \renewcommand{\arraystretch}{1.05}
  \caption{Guarantees codebook from axial coding.}
  \vspace{-2.5mm}
  \label{tab:guarantee-tags}
  \begin{tabular}{p{0.3\linewidth} p{0.66\linewidth}}
    \toprule
    \textbf{(Tag) Name} & \textbf{Meaning} \\
    \midrule
    (G\_SAFE) Safety & Stays within a safe set; no constraint violations. \\
    (G\_STAB) Stability & States remain bounded or converge w/o disturbances. \\
    (G\_PERF) Performance & Meets accuracy, efficiency, or cost objectives. \\
    (G\_FEAS) Feasibility & Valid solutions/trajectories exist under constraints. \\
    (G\_TIME) Timing & Meets computation/comm. deadlines and jitter bounds. \\
    (G\_ROBUST) Robustness & Guarantees persist under bounded uncertainty, modeling error, and disturbances. \\
    (G\_RELY) Reliability & Maintains guarantees with high probability despite random failures. \\
    (G\_SECURE) Security & Preserves integrity, confidentiality, and availability (CIA) against adversaries. \\
    (G\_RESIL) Resilience & Recovers from faults; degrades gracefully while maintaining function. \\
    \bottomrule
  \end{tabular}
\end{table}

\Q{How were preliminary categories defined and refined?}

\subsubsection{Selective Coding: Categorizing Assumptions Tags}
Starting from the axial assumption tag inventory, we performed selective coding on assumptions to integrate tags into higher-level categories and to organize the assumption-guarantee mapping used in analysis (guarantee tags were not further categorized). We enforced a single-parent policy at the tag level (each tag belongs to exactly one high-level category), while allowing entries (rows) to carry multiple tags across categories. The scheme was iteratively refined until (i) new papers mapped cleanly without creating additional categories and (ii) re-reads of earlier papers did not trigger systematic reassignments. Disagreements were resolved by adjudication, and a 10–20\% sample was double-coded to check consistency.

The resulting high-level categories in Tab.~\ref{tab:assumption-categories} are used to aggregate counts, structure the assumption-guarantee co-occurrence analysis, and support the high-level observations in Sec.~\ref{sec:high-level-observations}, while the underlying per-row tags (Tab.~\ref{tab:assumption-tags}) are used for tag-level plots (Figs.~\ref{fig:assumption_tags_stacked_bar_percent}--~\ref{fig:assumption_tags_stacked_bar}) and detailed analyses.

\begin{table}[h]
  \centering
  \small
  \caption{Broad assumption categories from selective coding.}
  \vspace{-2.5mm}
  \label{tab:assumption-categories}
  \begin{tabular}{p{0.34\linewidth} p{0.56\linewidth}}
    \toprule
    \textbf{Category} & \textbf{Included assumption tags} \\
    \midrule
    Modeling & A\_ABS, A\_MVAL, A\_NN, A\_ARCH \\
    External Constraints & A\_ENV, A\_HUM, A\_FAULT, A\_UTIL \\
    Physical & A\_PHY, A\_ACTU, A\_INIT, A\_TIME \\
    Interface and Estimation & A\_SENS, A\_PERCEP \\
    \bottomrule
  \end{tabular}
\end{table}

\section{Summary of Corpus}
\label{sec:summary-of-literature-data}

This section characterizes how current CPS work uses assumptions, guarantees, and formal language elements. We first quantify which assumption categories, tags, and language features appear most often, revealing a strong skew toward modeling assumptions expressed with algebraic and logical features. We then summarize the distribution of guarantees and analyze assumption–guarantee co-occurrences to reveal systematic gaps.

\Q{Overall, what assumptions show up the most?}
\subsection{Distributions of Assumptions}
\begin{figure}
    \centering
    \includegraphics[width=1\linewidth]{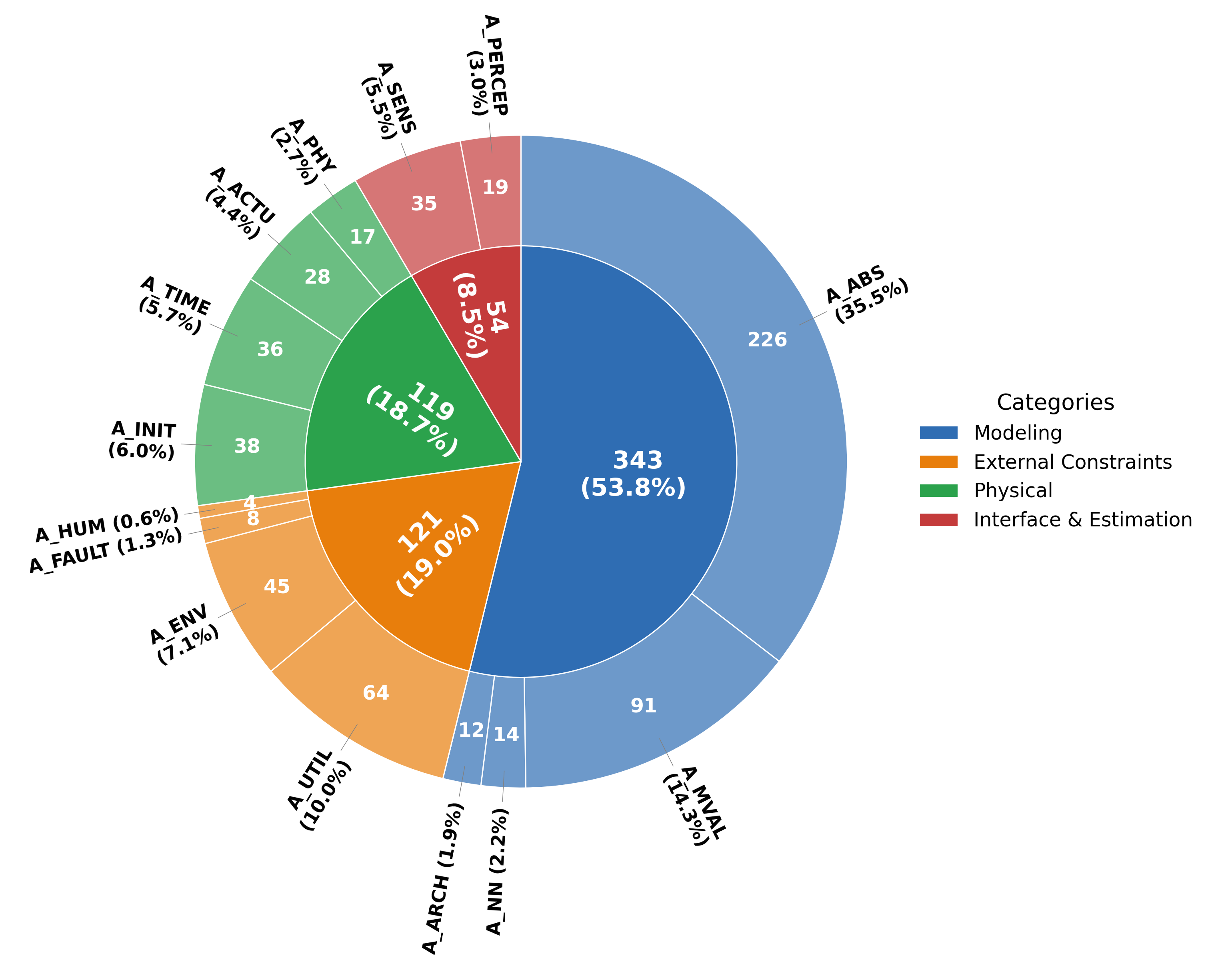}
    \caption{Distribution of assumption tags and categories across the corpus ($N{=}637$). The inner ring shows the share of each high-level category; the outer ring breaks each category into its constituent tags.}
    \label{fig:pie_assumptions}
\end{figure}

At the assumption category level, modeling (53.8\%) dominates as the most common type of assumptions, followed by external constraints (19.0\%), physical (18.7\%), and interface and estimation (8.5\%). Fig.~\ref{fig:pie_assumptions} summarizes these category-level shares across the corpus ($N{=}637$). This highlights the central role of models in providing CPS guarantees.

At the assumption tag level, the distribution is similarly biased towards model-related aspects. Fig.~\ref{fig:pie_assumptions} shows that system abstraction (\texttt{A\_ABS}) is the most common tag, representing 35.5\% of all assumptions, followed by model validity (\texttt{A\_MVAL}) at 14.3\%. Together, these two tags account for nearly half of all instances. Three additional tags --- utilities and priors (\texttt{A\_UTIL}, 10.0\%), environment and agents ($\texttt{A\_ENV}$, 7.1\%), and state envelope ($\texttt{A\_INIT}$, 6.0\%) --- contribute another quarter, bringing the top five tags to approximately 73\% of the corpus.

The remaining assumption tags are less represented, each contributing less than 6\%. These include timing and scheduling ($\texttt{A\_TIME}$), sensing ($\texttt{A\_SENS}$), perception (\texttt{A\_PERCEP}), actuation (\texttt{A\_ACTU}), physical dynamics (\texttt{A\_PHY}), neural regularity (\texttt{A\_NN}), neural architecture (\texttt{A\_ARCH}), fault and tamper (\texttt{A\_FAULT}), and human behavior (\texttt{A\_HUM}). However, some of these aspects, particularly timing and perception, are occasionally and implicitly represented inside larger abstract models (annotated by $\texttt{A\_ABS}$).

\looseness=-1
Overall, the distribution of assumptions indicates that most CPS studies rely on modeling assumptions about how systems are abstracted, validated, or parameterized, whereas explicit assumptions about timing, sensing, and human behavior are much less common.

\Q{Overall, what guarantees show up the most?}
\begin{figure}
    \centering    \includegraphics[width=0.8\linewidth]{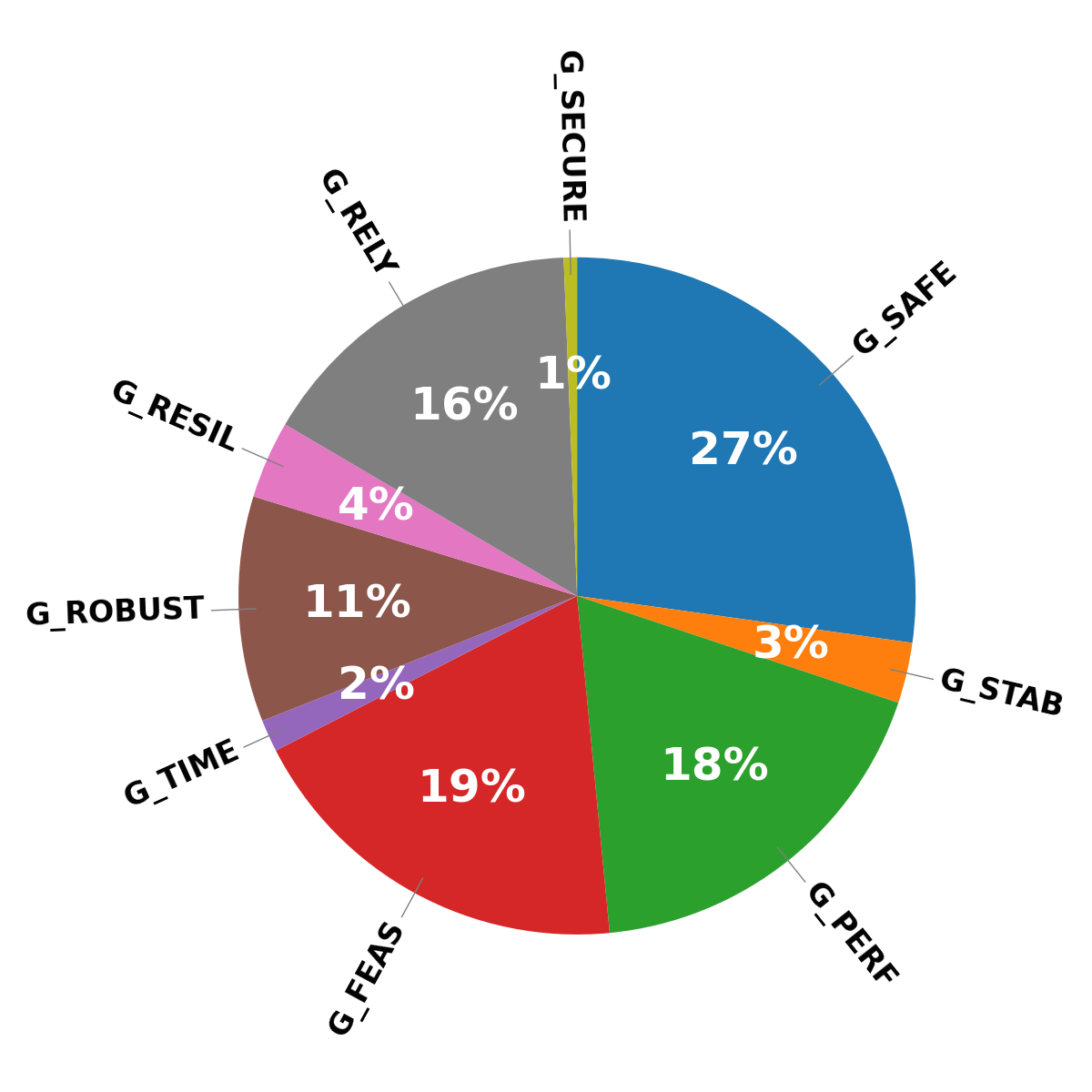}
    \vspace{-6mm}
    \caption{Distribution of guarantee tags across the corpus ($N{=}496$).}
    \label{fig:pie_overall_guarantees}
\end{figure}

\subsection{Distribution of Guarantees}

Fig.~\ref{fig:pie_overall_guarantees} summarizes the shares of guarantee tags across all contexts in the corpus ($N=496$). The distribution is dominated by safety (\texttt{G\_SAFE}, $27\%$). Three mid-sized categories follow: feasibility (\texttt{G\_FEAS}, $19\%$), performance (\texttt{G\_PERF}, $18\%$), and reliability (\texttt{G\_RELY}, $16\%$). Together, these four account for nearly $80\%$ of all guarantee statements. The remaining tags form a long tail of less frequent properties: robustness (\texttt{G\_ROBUST}, $11\%$), resilience (\texttt{G\_RESIL}, $4\%$), stability (\texttt{G\_STAB}, $3\%$), timing (\texttt{G\_TIME}, $2\%$), and security (\texttt{G\_SECURE}, $1\%$). In part, these low counts are due to our survey not specifically focusing on the security and timing sub-areas of the CPS literature.  

\looseness=-1
Overall, safety dominates as the primary sought-after guarantee, followed by feasibility, performance, and reliability. Robustness appears moderately often, while resilience and stability guarantees remain relatively rare. This pattern indicates that most CPS work still focuses on ensuring the desired properties under nominal conditions, rather than perturbed, adversarial, or post-failure contexts.

\Q{Overall, what language features show up the most?}
\subsection{Language Features for Assumptions}

\begin{figure}
    \centering
    \includegraphics[width=1\linewidth]{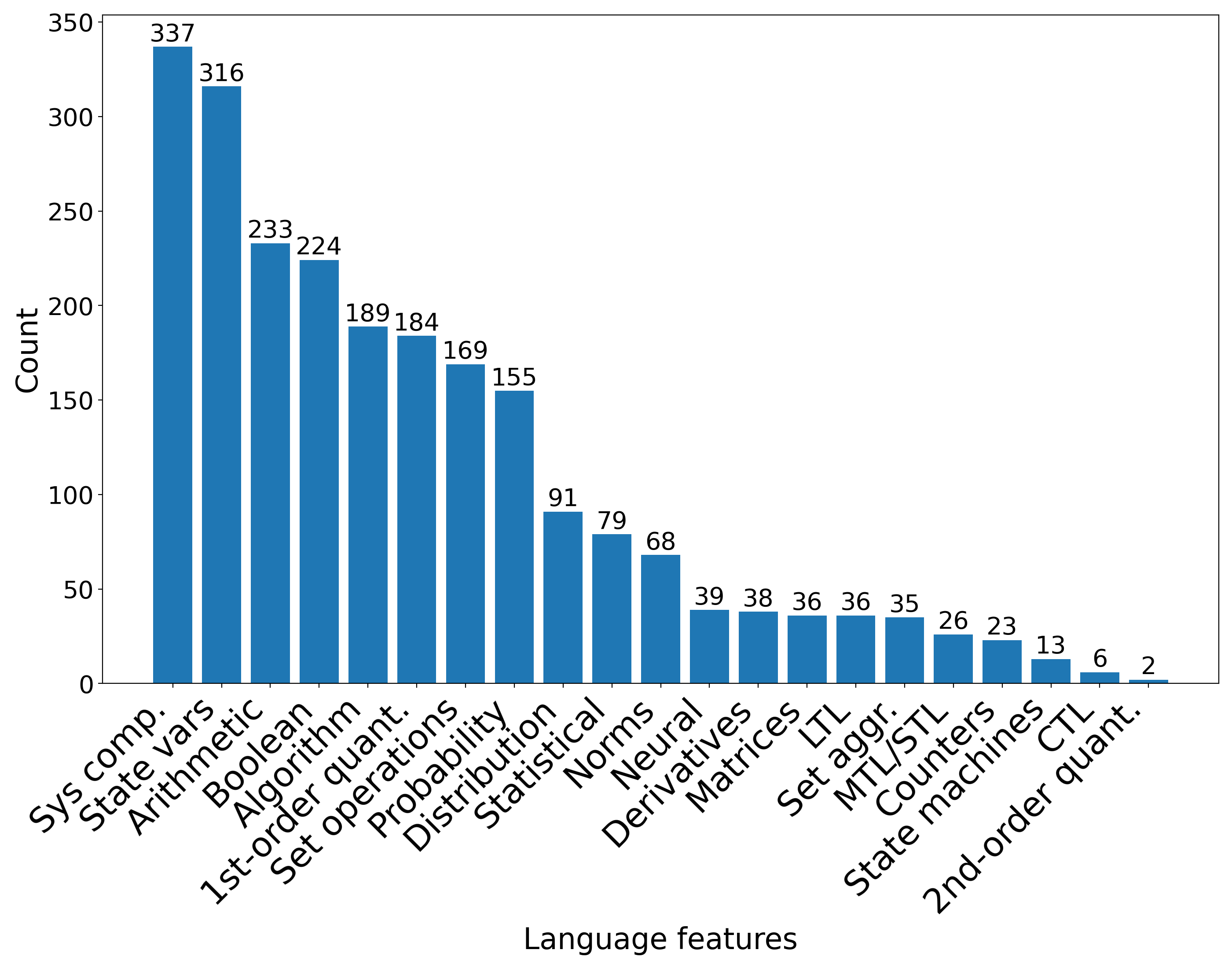}
    \vspace{-6mm}
    \caption{Counts of language features used to formalize assumptions ($N{=}2{,}299$ feature instances).}
    \vspace{-3mm}
    \label{fig:lang_features_bar}
\end{figure}

\looseness=-1
Fig.~\ref{fig:lang_features_bar} summarizes which language features were used by assumptions across the corpus ($N{=}2{,}299$). Algebraic and logical primitives, along with variables and components, are common to nearly all assumptions, providing the basic tools for CPS-relevant assertions. That is, the most common features are \texttt{system components} (14.7\%) and \texttt{state variables} (13.7\%), followed by \texttt{arithmetic} (10.1\%) and \texttt{boolean} (9.7\%). Next most common features are \texttt{algorithm} (8.2\%), \texttt{first-order quantifiers} (8.0\%), and \texttt{set operations} (7.4\%). Here, the \texttt{algorithm} feature typically marks explicit computational procedures, such as a model-predictive control optimization algorithm or an SMT-based verification back-end algorithm that searches for counterexamples to a temporal-logic specification. Together, these eight features account for 78.6\% of all feature instances.

\looseness=-1
Uncertainty-related features appear fairly often as well. Together, \texttt{probability} (6.7\%), \texttt{distribution} (4.0\%), and \texttt{statistical} (3.4\%) account for 14.1\% of the corpus. Next, a smaller group of geometric and numerical features supports assumptions involving quantities, magnitudes, or model structure, including \texttt{norms} (3.0\%), \texttt{derivatives} (1.7\%), \texttt{matrices} (1.6\%), and \texttt{neural} (1.7\%). Finally, the remaining features, including temporal and branching logic, are rarely used. Across all papers, \texttt{LTL} (1.6\%), \texttt{MTL/STL} (1.1\%), and \texttt{CTL} (0.3\%) together make up only 3.0\%, whereas \texttt{state machines} (0.6\%) and \texttt{second-order quantifiers} (0.1\%) are almost never observed.

\Q{Empirical Gaps about Assumptions}
\subsection{Assumption–Guarantee Co‑occurrence}
\begin{figure}
    \centering
    \includegraphics[width=\linewidth]{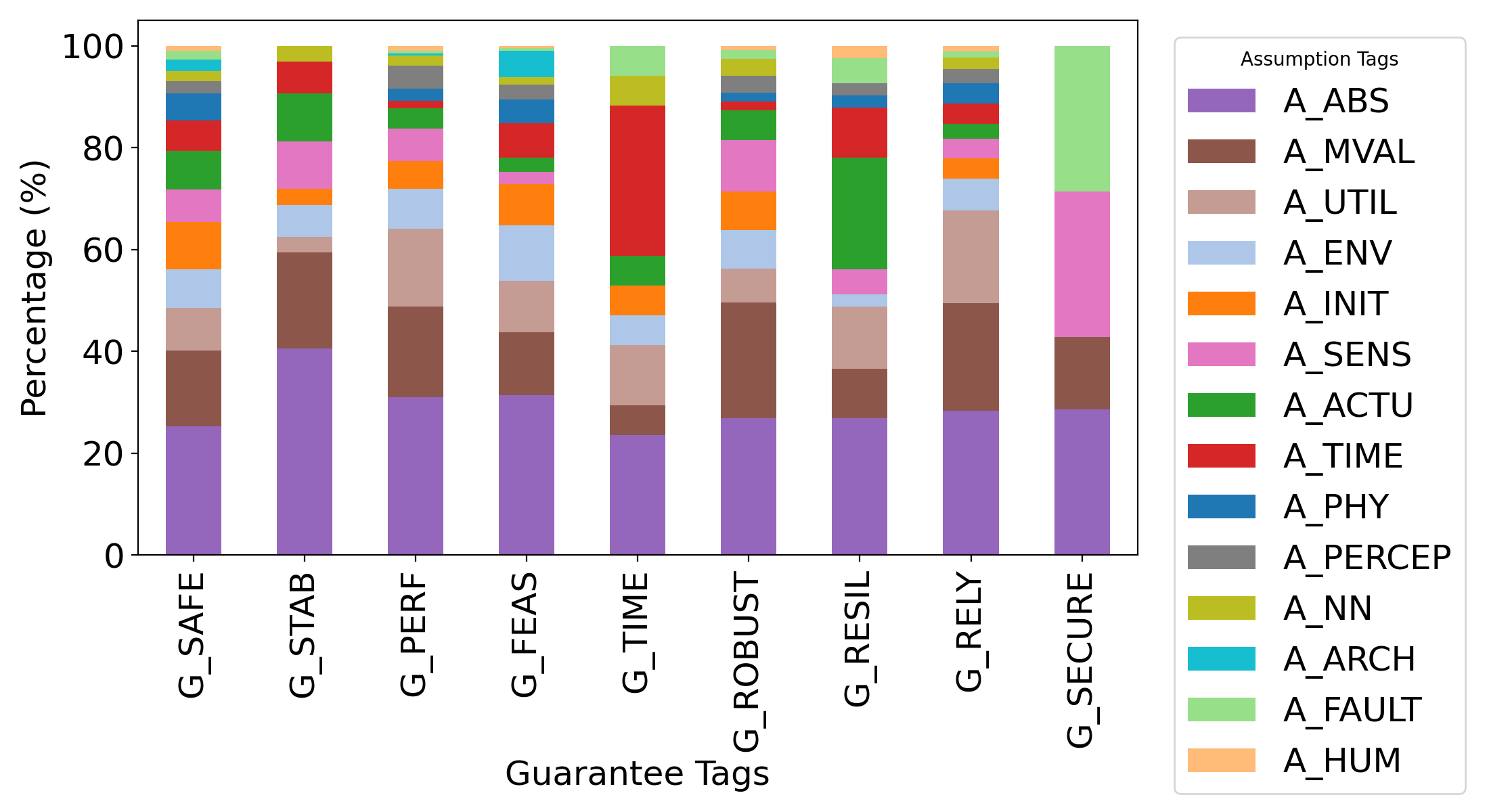}
    \caption{Percent-normalized co-occurrence of assumption tags (colors) with guarantee tags (x-axis). Each bar sums to $100\%$ and shows the internal composition of assumptions supporting a given guarantee.}
    \label{fig:assumption_tags_stacked_bar_percent}
\end{figure}

\begin{figure}
    \centering
    \includegraphics[width=1\linewidth]{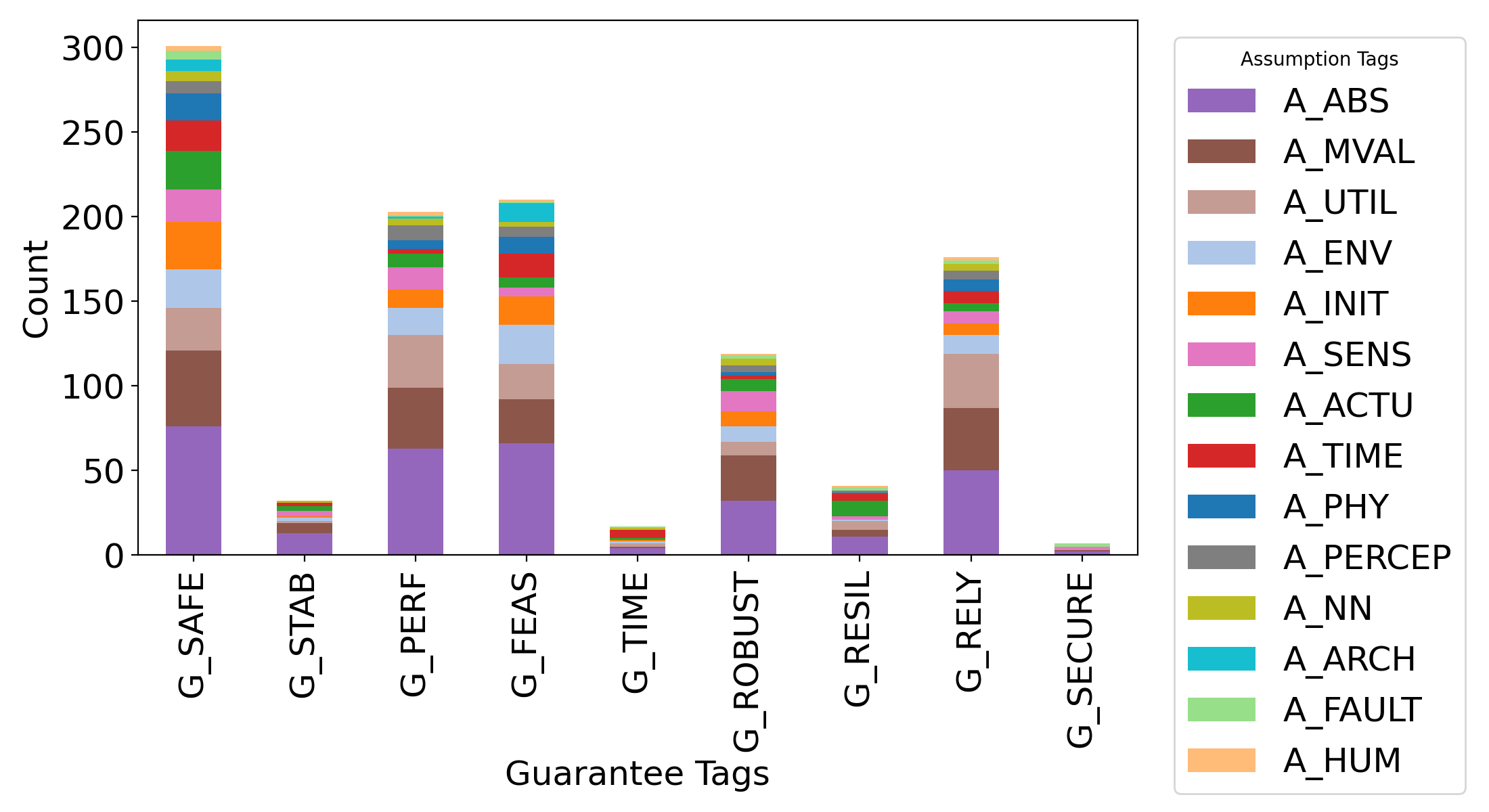}
    \caption{Count-normalized co-occurrence of assumption tags with guarantee tags. Bar heights show the total number of assumption instances attached to each guarantee.}
    \label{fig:assumption_tags_stacked_bar}
\end{figure}

Our next step was to examine how often each assumption tag co-occurs with each guarantee by analyzing the percent- and count-normalized cross-tag plots. The percent view in Fig.~\ref{fig:assumption_tags_stacked_bar_percent} highlights the internal composition of assumptions within a given guarantee. Simultaneously, the count view in Fig.~\ref{fig:assumption_tags_stacked_bar} confirms that these proportions reflect real volume rather than small-sample artifacts. Together, they reveal clear structural regularities: \emph{modeling} assumptions (\texttt{A\_ABS}, \texttt{A\_MVAL}, \texttt{A\_NN}, \texttt{A\_ARCH}) dominate across all guarantees, whereas \emph{interface and estimation} assumptions, including perception and sensing (\texttt{A\_PERCEP}, \texttt{A\_SENS}), appear less frequently than expected in the contexts where they should be most relevant (e.g., safety, feasibility, and performance). Conversely, a few assumption types, such as initialization (\texttt{A\_INIT}), environment (\texttt{A\_ENV}), and fault modeling (\texttt{A\_FAULT}), show higher-than-expected concentrations under specific guarantees, indicating localized dependencies rather than balanced coverage across the corpus. We summarize the results out of our analysis below. 

\paragraph{Modeling Assumptions Dominate across Guarantees (\texttt{A\_ABS}, \texttt{A\_MVAL})}
Across nearly all guarantee columns in Figs.~\ref{fig:assumption_tags_stacked_bar_percent}--\ref{fig:assumption_tags_stacked_bar}, system abstraction (\texttt{A\_ABS}) provides the largest share and model validity (\texttt{A\_MVAL}) the second, indicating that guarantees of every type are primarily justified by how the system is abstracted and how its model is trusted rather than by environment, timing, or component-specific factors. Once again, the evidence confirms the central role of abstract models in providing rigorous guarantees.

\paragraph{Guarantees Conditioned on Utility Priors (\texttt{A\_UTIL})}
Utility assumptions appear at moderate but visible levels across multiple guarantees in Figs.~\ref{fig:assumption_tags_stacked_bar_percent}--\ref{fig:assumption_tags_stacked_bar}. These assumptions complement the modeling tags rather than replace them. They capture reliance on external fixed resources and priors such as black-box simulators, fixed risk or cost maps, differentiable renderers, or pre-trained embeddings. This pattern reveals that many guarantees are conditional not only on the correctness of the model but also on the correctness of the surrounding utility components. When simulators, risk maps, or pre-trained features are misaligned with deployment, the stated guarantees may fail even if the underlying dynamics model remains accurate.

\paragraph{Underspecified Perception Assumptions (\texttt{A\_PERCEP})}
Since perception errors such as missed obstacles, misclassified lanes, or outdated detections can play a significant role in unsafe CPS behavior~\cite{rahman_run-time_2021, jansen_formal_2025, gros_architectural_2022}, one might expect perception assumptions (\texttt{A\_PERCEP}) to appear frequently as supporting conditions for safety guarantees (\texttt{G\_SAFE}). However, perception-related assumptions occur less often in safety guarantees and appear more regularly in feasibility and performance contexts. In both the percent- and count-normalized plots (Figs.~\ref{fig:assumption_tags_stacked_bar_percent} and \ref{fig:assumption_tags_stacked_bar}), the perception (\texttt{A\_PERCEP}) segment within safety (\texttt{G\_SAFE}) occupies a smaller share than in feasibility (\texttt{G\_FEAS}) or performance (\texttt{G\_PERF}), and remains limited compared with the dominant modeling tags system abstraction (\texttt{A\_ABS}) and model validity (\texttt{A\_MVAL}). This pattern points to a shortage of explicit perception assumptions underlying safety, suggesting that the connection between perception reliability and safety outcomes is often left implicit or underspecified. We zoom in on this observation in the next section.

\paragraph{Underspecified Sensing Assumptions (\texttt{A\_SENS})} 
In the percent- and count-normalized plots (Figs.~\ref{fig:assumption_tags_stacked_bar_percent}--\ref{fig:assumption_tags_stacked_bar}), \texttt{A\_SENS} occupies only a modest share across most guarantee columns compared with the dominant modeling tags (\texttt{A\_ABS}, \texttt{A\_MVAL}). This pattern holds even when the guarantee directly depends on measurement quality --- such as feasibility under estimation constraints or safety under sensor dropouts --- where one would normally expect \texttt{A\_SENS} to be prominent. Instead, within \texttt{G\_SAFE} and \texttt{G\_FEAS}, its portion is smaller than initialization and utility and remains well below the dominant modeling assumptions. This suggests that many analyses implicitly assume accurate and reliable sensing without stating it explicitly, leaving the sensing layer underspecified even in contexts where observability or sensor performance is critical to the claimed guarantee. For example, Yang~\cite{yang_safety_2021} makes the sensing assumption fully explicit by requiring that the measurement-loss indicator sequence satisfy an automaton-constrained pattern such as a $(m,k)$-firm bound, where in every window of length $m$ at most $k$ measurements are missing; the safety guarantee is proven only under this bounded-loss sensing condition.

\looseness=-1
\paragraph{Rare Explicit Neural Assumptions (\texttt{A\_NN})}  
Given the widespread use of learned perception, estimation, and control components in CPS, one might expect \texttt{A\_NN} to appear frequently across guarantees. In practice, neural-related assumptions are rare in the corpus. Both the percent- and count-normalized plots (Figs.~\ref{fig:assumption_tags_stacked_bar_percent} and \ref{fig:assumption_tags_stacked_bar}) show that \texttt{A\_NN} occupies only a minor share in every guarantee column, including \texttt{G\_PERF} and \texttt{G\_SAFE}, where neural modules are common in implementation. In the few cases where \texttt{A\_NN} appears, the assumptions typically describe training-time properties such as regularization or bounded Lipschitz continuity. Most guarantees that include neural components are instead tagged with \texttt{A\_ABS} or \texttt{A\_MVAL}, suggesting that neural networks are generally treated indirectly as instances of the broader types of model (e.g., smooth functions) --- rather than as components with unique structure and behavior.

\paragraph{Initialization-dependent Safety and Robustness (\texttt{A\_INIT})}
Initialization is expected to be a major assumption for feasibility, since initialization conditions define whether valid trajectories or reachable sets exist, while they should play only a minor role for most other guarantee types. By contrast, stability is typically treated as largely initialization-independent; for example, Lyapunov-region arguments aim to hold over a fixed domain that is not tuned to a particular starting state. Indeed, across the percent- and count-normalized plots Figs.~\ref{fig:assumption_tags_stacked_bar_percent} and \ref{fig:assumption_tags_stacked_bar}), state envelope assumption (\texttt{A\_INIT}) takes a fair share of feasibility (\texttt{G\_FEAS}) but only a marginal share of stability (\texttt{G\_STAB}), matching this expectation. However, it also contributes a non-trivial portion of safety (\texttt{G\_SAFE}) and robustness (\texttt{G\_ROBUST}), suggesting that many of these guarantees hold only within declared starting envelopes, rather than being global, revealing a gap in global robustness and safety guarantees. We highlight this observation in the next section.

\section{High-level Observations}
\label{sec:high-level-observations}

This section interprets the corpus-level patterns from Sec.~\ref{sec:summary-of-literature-data} and highlights four cross-cutting gaps in current CPS practice. Sec.~\ref{subsec:init-dependent-safety-robustness} shows that many safety and robustness guarantees remain tied to particular initial-state regions rather than holding globally. Sec.~\ref{subsec:limited-perception-sensing} argues that perception and sensing conditions are rarely stated, even when guarantees depend critically on information quality. Sec.~\ref{subsec:limited-neural-assumptions} examines how neural components are typically hidden inside generic modeling assumptions instead of being accompanied by neural-specific conditions. Finally, Sec.~\ref{subsec:gaps-uncertainty} discusses how uncertainty is often encoded indirectly through sets and algebra rather than via an explicit probabilistic language, complicating comparison and transfer of guarantees across deployments.

\subsection{Initialization-Dependent Safety/Robustness}
\label{subsec:init-dependent-safety-robustness}

Many safety and robustness guarantees remain initialization-dependent. Feasibility is naturally tied to initial conditions, whereas closed-loop properties such as stability, safety, and robustness are ideally stated to hold uniformly over an operating envelope, largely independent of where the system starts. 

Figs.~\ref{fig:assumption_tags_stacked_bar_percent}--\ref{fig:assumption_tags_stacked_bar} show the expected prominence of state-envelope assumptions (\texttt{A\_INIT}) for feasibility guarantees (\texttt{G\_FEAS}), but also a notable share for safety (\texttt{G\_SAFE}) and robustness (\texttt{G\_ROBUST}), while stability (\texttt{G\_STAB}) remains small. This indicates that many safety and robustness claims in the literature are restricted to particular initial-state regions, with guarantees that hold only within declared starting envelopes rather than globally. Such local guarantees are not necessarily invalid, but they complicate comparison and reuse-results may not transfer when the initial-state distribution changes, two ``robust'' methods can cover disjoint operational regions, and downstream users may assume global properties that were never demonstrated. 

\subsection{Limited Perception/Sensing Assumptions}
\label{subsec:limited-perception-sensing}

Perception and sensing assumptions are underspecified in guarantees that rely on accurate measurement and reliable interpretation of the environment. The cross-tab results (Figs.~\ref{fig:assumption_tags_stacked_bar_percent}--\ref{fig:assumption_tags_stacked_bar}) indicate that perception (\texttt{A\_PERCEP}) contributes only a small portion within safety (\texttt{G\_SAFE}), while sensing (\texttt{A\_SENS}) remains limited across both feasibility (\texttt{G\_FEAS}) and safety; in contrast, modeling assumptions (\texttt{A\_ABS}, \texttt{A\_MVAL}) dominate these guarantees.

\looseness=-1
This stands in contrast to the central role that perceptual information plays in CPS practice. Feasibility and safety often depend on what can be reliably measured and accurately interpreted-factors such as coverage, latency, noise, dropouts, and detection or recall directly affect whether safe or feasible behavior is even observable or enforceable. When perception and sensing assumptions are left implicit, guarantees become brittle under shifting conditions, difficult to reproduce across setups with varying sensor configurations, and prone to misinterpretation-particularly when performance depends more on information quality than on the control architecture itself.

\subsection{Limited Neural Assumptions}
\label{subsec:limited-neural-assumptions}

Although neural networks have become a predominant way to solve data-driven CPS tasks, assumptions on neural architectures and regularity are rarely stated explicitly across guarantees. Neural assumptions (\texttt{A\_NN}) appear only sparsely across all guarantees (Sec.~\ref{sec:summary-of-literature-data}). Despite the prevalence of learned components in perception, estimation, and control, most papers fold them into broader modeling (\texttt{A\_ABS}, \texttt{A\_MVAL}) or design (\texttt{A\_UTIL}) assumptions rather than stating their neural properties explicitly.

\looseness=-1
This lack of neural-specific assumptions creates an attribution gap: when a guarantee holds or fails, it is unclear whether the neural component’s properties were examined, relevant, or responsible. The result is reduced comparability across studies (two controllers may both ``use a network’’ yet rely on very different, unstated behaviors) and weaker interpretability when failures occur. 

\subsection{Gaps in Expressing Uncertainty}
\label{subsec:gaps-uncertainty}
Many CPS guarantees depend on uncertainty arising from sensor noise, external disturbances, and dataset shifts, yet the literature rarely describes this uncertainty in a shared, explicit language. In Fig.~\ref{fig:lang_features_bar}, uncertainty-related terms (\texttt{probability}, \texttt{distribution}, \texttt{statistical}) account for only 14.1\% of mentions (6.7\%, 4.0\%, 3.4\%). In contrast, algebraic and first-order features such as \texttt{system components} (14.7\%), \texttt{state variables} (13.7\%), \texttt{arithmetic} (10.1\%), \texttt{boolean} (9.7\%), \texttt{first-order quantifiers} (8.0\%), and \texttt{set operations} (7.4\%) dominate the corpus. Notably, even feature families that appear less often overall, such as temporal and branching logics (LTL, MTL, CTL; 3.0\% combined), have clear and standardized notations. In comparison, uncertainty is frequently encoded indirectly through sets and algebra (e.g., worst‑case bounds) rather than through explicit probabilistic statements.

When uncertainty is represented only by fixed bounds, the degree of stochastic assurance behind a guarantee is often unclear (e.g., a chance constraint $\Pr[\text{collision}]\le\varepsilon$ versus a deterministic bound). This lack of clarity makes it difficult to compare guarantees across studies, reproduce results under new data or operating conditions, and limits the usefulness of such results for practitioners who must reason about residual uncertainty and distribution shift --- the contexts where robustness and resilience are expected to provide the most insight.
\section{Calls for action}
\label{sec:calls-for-action}

The empirical patterns identified in Section~\ref{sec:high-level-observations} reveal several structural gaps in how CPS assumptions are described and used. This section turns those observations into \textit{four concrete guidelines} for the CPS community on formulating, reporting, and benchmarking guarantees. We anticipate that following these guidelines would improve the reproducibility and comparability across studies --- and also facilitate the transfer of guarantees to real systems. 

\Q{Next Step for Initialization-Dependent Safety/Robustness}
\subsection{Analyze Initialization Dependence}
Many guarantees of safety and robustness depend on state initialization, which significantly impacts their application. To improve the clarity, replication, and composition of such guarantees, we recommend \textit{explicitly describing} each guarantee as initialization-dependent or independent, without altering the methodology. For initialization-dependent guarantees, one should report the \textit{geometry} and \textit{measure} of the initial set, along with a brief \textit{sensitivity analysis} to the set's size. Furthermore, future CPS benchmarks should vary initial-state regimes to reveal which claims are local versus global. 

\Q{Next Step for Limited Perception/Sensing Assumptions}
\subsection{Assert Sensing and Perception Constraints}
Guarantees that depend on sensor data frequently omit the underlying sensing and perception assumptions, making it difficult to understand the transfer to real-world systems (which vary in the noise level). To close this gap, we recommend that CPS researchers (i) state the conditions on the sensor/perception inputs that their guarantees rely on (e.g., bounds on sensor noise and dropouts, latency limits, minimum detection and recall for safety‑critical classes); (ii) explore the possibility that their guarantees extend to relaxed assumptions; (iii) include a brief sensitivity analysis showing how their results change as the assumptions on sensing/perception quality are tightened or relaxed. Those who develop benchmarks are encouraged to provide configurations with varied sensing and perception quality while holding the remaining parameters fixed.

\Q{Future Steps for Limited Neural Assumptions}
\subsection{Get Closer to Neural Assumptions}
Learning-enabled components are common --- yet neural assumptions are rarely stated in detail, leading to a gap with applications tied to specific neural architectures and properties. First, we ask researchers working on learning-enabled CPS to report minimal quantitative descriptors such as Lipschitz or sector bounds, calibration errors, or operating-domain limits. Reporting such assumptions does not require new analytic tools. Second, we recommend deepening the analysis of guarantees to link them more closely to the nature of the neural component in the loop.  Finally, on the benchmark side, we argue for including variants where only the neural component changes, revealing whether the guarantee is architecture-sensitive or architecture-agnostic.

\Q{Next Step for Gaps in Expressing Uncertainty}
\subsection{Report Uncertainty Systematically}
Rich, stochastic uncertainty is often simplified to non-determinism and fixed bounds, which tends to obscure risk levels and impacts of unexpected uncertainty. In addition, guarantees become hard to compare and reproduce across different uncertainty formulations. We recommend that CPS papers adopt a \textit{structured template} of reporting stochastic uncertainty in their guarantees, inspired by the Kolmogorov's probability spaces. First, identify the \textit{source(s) of randomness} in the disturbance/noise model; for instance, we tend to distinguish the offline data collection process $\mathcal{D}$ and the online chance trajectory distribution $\Omega$. Second, set the acceptable risk thresholds for the uncertainties; for instance, confidence $\delta$ for data collection $\mathcal{D}$ (e.g., if we repeated data collection $N$ times, for at least $(1-\delta)N$ of them, the guarantee would hold up) and chance constraint $\alpha$ on trajectories $\Omega$ (e.g., we sample $M$ trajectories, at least $(1-\alpha)M$ of them would uphold the guarantee). Finally, compose these stochastic uncertainties into a probably approximately correct (PAC)-style guarantee~\cite{kearns_introduction_1994}: 
$$  \Pr\nolimits_\mathcal{D}\Big[  \Pr\nolimits_\Omega[\texttt{guarantee} ] \ge 1-\alpha \Big] \ge 1-\delta  $$

Naturally, we also recommend a brief sensitivity analysis between the identified randomness and the strengths of the guarantees. The experimental validation of uncertainty-related effects needs to follow methodological rigor (for example, confidence intervals are prone to frequent misinterpretations~\cite{greenland_statistical_2016}).
\section{Limitations}
\label{sec:limitations}

Here we critically examine the limitations of our survey along with their mitigations and effects on interpreting the results.

\subsubsection*{Construct Validity: Tags and Taxonomies}
The central constructs of assumptions and guarantees were defined formally and accompanied by scoping remarks, to minimize the risk of misinterpretation. However, since manual coding is inherently subjective and terminology can vary, our tag taxonomies have the risk of missing nuances of overlapping concepts. In particular, broad tags, such as \texttt{A\_ABS} (abstraction), risk becoming a catch-all category that absorbs instances better suited to narrower tags (e.g., \texttt{A\_SENS}, \texttt{A\_PERCEP}, \texttt{A\_NN}). Similarly, boundaries between related tags like \texttt{A\_ENV}, \texttt{A\_HUM}, \texttt{A\_PERCEP}, and \texttt{A\_SENS} can be ambiguous. We mitigated these ambiguities by cross-coding a small subset of studies for reliability and holding regular discussions across the group of authors. Future surveys could make the coding scheme more robust by publishing the codebook with positive and negative examples. 

\looseness=-1
\subsubsection*{Internal Validity: Quantification, Analysis, and Observations}
This is an observational meta-analysis focused on published assumptions and guarantees --- not an interventional or experimental study. Our quantitative results should be interpreted as \textit{descriptive} rather than predictive or causal. To illustrate varying strengths of evidence, our figures report frequencies using count- and percent-normalized plots. Consequently, rare cells may appear disproportionately large, so the readers should take care not to interpret low-sample proportions or comparisons as evidence of statistical association. To improve transparency and correct any errors that threaten the internal validity of our study, we publish our 
\href{https://docs.google.com/spreadsheets/d/e/2PACX-1vRd6uMD7AQFMFZUouHEKP923pCrlstxin7xkRuTwRGkfXJ2S6dByGgaXCpOUNWd4UAqsv1UB9fU0kvH/pubhtml}{\textbf{tagged database}}, and provide an \href{https://docs.google.com/forms/d/e/1FAIpQLSfuPPUTjk_LMuJDBtHnVeN2szL1k7KP9pccBik7Q14lMJ4ksw/viewform?usp=dialog}{\textbf{online form}} where readers can suggest additional papers or corrections. Future versions of the dataset will incorporate vetted submissions to address coverage and citation gaps.

\subsubsection*{External Validity: Scope, Corpus, and Generalizability.}
Our corpus is a snapshot of recent CPS literature and does not exhaustively represent sub-domains, venues, or years. In particular, several guarantee tags contain relatively few papers (e.g., timing, security, stability), so apparent differences in those columns may reflect sampling variability rather than systematic priorities. By limiting the corpus to a fixed set of search keywords and explicit guarantees, our sample likely over-represents modeling and verification while under-representing perception or deployment case studies. Thus, our conclusions may not generalize to significantly different distributions over CPS papers --- or to the CPS practice. The identified gaps may not represent a true absence in deployed systems. Instead, these omissions could stem from scholarly writing conventions or publication constraints, such as page limits. 
Future surveys that seek to strengthen generalization should set quotas by domain and guarantee type, formalize inclusion and exclusion criteria and search strings, and maintain an updated corpus that is regularly revised, to avoid dependency on a single cohort of papers. In addition, future research should replicate these findings within specific domains (e.g., autonomous driving and industrial automation) through expert interviews, artifact mining, and participatory case studies.
\section{Conclusion}
\label{sec:conclusion}

This survey aimed to identify assumptions and guarantees in cyber-physical systems (CPS). Through a systematic analysis of 104 papers, we tagged 423 assumptions, 321 guarantees, and 2,299 instances of language features. We mapped the expression of assumptions, the guarantees they support, and their distribution between offline and online contexts. The resulting codebook, dataset, and taxonomy are intended to serve as a reference for specification.

The analysis revealed \textit{several clear empirical signals}. First, modeling assumptions, particularly those concerning system abstraction and model validity, dominate across all guarantees. Second, interface-level requirements for sensing, perception, and neural components are comparatively rare, even in domains where they are most critical. Third, many safety and robustness claims remain explicitly initialization-dependent, with guarantees holding only within declared starting envelopes rather than globally. Fourth, neural-specific assumptions appear only sparsely and are often folded into broader modeling or utility tags, making it difficult to attribute the role of learned components. Finally, the formalization of assumptions relies heavily on algebraic and first-order features, whereas probabilistic or temporal structures and explicit neural predicates are seldom used, leading to ambiguity about the level of stochastic assurance behind guarantees.

This study provides a descriptive overview rather than an exhaustive survey. The scope is defined by several limitations: the corpus is concentrated on CPS research with explicit guarantees, the coding was performed manually, and the analysis reports frequencies rather than inferential statistics. Future work should include replication in specific subdomains and impact-weighted analyses. Despite these limitations, the observed regularities and gaps provide an informative basis for enhanced reporting practices and targeted benchmarks.

Together, the taxonomy and artifacts from this study are offered as a common framework for naming, validating, and relating assumptions to guarantees. Adopting clearer assumption reporting and minimal sensitivity analyses can substantially improve the reproducibility and comparability of CPS results and better support the transfer of formal guarantees to deployed systems.

\begin{acks}
This work was supported by the NSF CAREER Grant CNS 2440920. Any opinions, findings, or conclusions expressed in this material are those of the authors and do not necessarily reflect the views of the National Science Foundation (NSF) or the U.S. Government.
\end{acks}

\bibliographystyle{ACM-Reference-Format}
\bibliography{references}

\end{document}